# ROLE OF INFORMATION AND ICTS AS DETERMINANTS OF FARMER'S ADAPTIVE CAPACITY TO CLIMATE RISK: AN EMPIRICAL STUDY FROM HARYANA, INDIA

Priya Chetri, School of Public Policy IIT Delhi, chetripriya94@gmail.com

Upasna Sharma, School of Public Policy IIT Delhi, upasna@iitd.ac.in

P. Vigneswara Ilavarasan, Department of Management Studies IIT Delhi, vignes@iitd.ac.in

**Abstract:** Using the primary data collected for 463 farmers in six districts of Haryana, India, the present study attempts to understand the constituents of farmer's adaptive capacity at local level and how it can be enhanced. We use path analysis technique using the lavaan package in RStudio to empirically test the role of information. We find that information is a direct and significant contributor to enhancing farmers' adaptive capacity. However, even with exponential growth in use of technology, particularly information and communication technologies (ICTs), small farmers still lack access to information which hinders their capacity to respond to weather and climate risks. Thus, understanding the mechanism that can facilitate exchange and use of information by the farming community more effectively is important. We take an ensemble view of ICTs operationalized using ICT ecosystem and find significant interlinkages between information, technology and the ICT ecosystem that facilitate learning and information exchange and therefore contribute to enhancing farmers' adaptive capacity and building resilience to climate shocks. We find that ICT ecosystem does facilitate access to information and also mediate the effect of farmer's capability and willingness to use ICTs for agricultural purposes. Development of sound ICT ecosystem is likely to help farmers to better respond to changing climate in the future.

## 1. INTRODUCTION

About one-third of global crop yield variability of major crops like rice, wheat, maize is explained by variation in climate (Ray et al., 2015). Agricultural economies like India are expected to be highly impacted due to the increased variability in extreme weather events (Hoegh-Guldberg et al., 2018). Being resilient to such shocks is imperative to moderate the impact of such variability. Farm level resilience can be understood as the ability of farms to adapt to climatic, social, and market risk (Meuwissen et al., 2019). Enhancing adaptive capacity of an exposed unit is an important strategy for building resilience. Resources such as natural resources, financial and economic resources, human resources, technology, and information are considered significant contributors to build adaptive capacity (Brooks & Adger, 2005; Dillow, 2008; Jones et al., 2010). However, in the short time-span it is very difficult to alter the endowment set of some of these resources to enhance adaptive capacity. With a given endowment of resources, adequate information on risks and vulnerabilities helps in identifying adaptation needs[1] and options[2].

---

[1] Adaptation needs refer to circumstances requiring information, resources, and action to ensure safety of population and surety of assets in response to climate impacts. (IPCC AR5, pp. 839)

[2] Adaptation options are defined as the array of strategies and measures available and appropriate to address adaptation needs (IPCC AR5, pp. 840).

Access to information is an important determinant of farmer's adaptation strategies to climate variability both as a response to short term shocks and long term response to climate stressors (Alemayehu & Bewket, 2017; Deressa et al., 2009). Yet small farmers have limited access to information (FAO, 2019). Traditional sources of information such as extension services have limited outreach whereas information from local input vendors face issues related to reliability. Though ICTs have much broader outreach and are cost effective in terms of information dissemination, anecdotal evidences suggest that they do not ensure access to information. On the face of it, information seems accessible. However, constraints like mismatch in timing of TV/radio broadcast and farmer's working hour, not carrying phones in the field to avoid losing them while working hinder farmer's access to relevant information.

Thus, it is not just access to an underlying set of technologies (ICTs for example) but also the techniques for dissemination and exchange of information which are important to ensure information access to farmers. Various studies have demonstrated that even when individual farmer does not have access to modern ICTs, being connected to a social network where such technologies are accessed fill this space. Nesheim et al. (2017) found that farmers had less interest in accessing agri-met information themselves as they already receive it through their social networks. This entails the importance of local groups, neighbours, friends, relatives in serving the information needs of farmers (Kalusopa, 2005).

It helps in understanding how social networks and institutions interact with technology to facilitate the coping and adaption strategies of farmers to deal with the weather-climate uncertainty. ICT ecosystem is one such set of formal and informal institutions and networks that may influence the access to and use of technologies for dissemination, exchange, and use of information. Through spill-over mechanism yielded through social interactions, peer-effects, externalities and other types of interferences (Vazquez-Bare, 2017), ICT ecosystem can endorse technology and information use among farmers.

If adaptive capacity is about responding to risk better, then ICT ecosystems are particularly important to be studied and examined in the scholarship of any adaptation intervention (e.g., IMD's agri-met advisories which provides information to farmers to respond to risk better). In this study we attempt to explore and unpack the inter-relationship between information (climate information in particular), technology (ICTs in particular) and institutions (farmer's ICT ecosystems) and the effect that these have on the farmer's adaptive capacity to take adaptation action in response to climate risk. In the next section we discuss the relevant literature and the theoretical framework to conceptualize farmer's adaptive capacity at a local scale. The third section presents the methodology adopted in collection of data and operationalization of the variables. Section 4 presents the main findings of the statistical analysis. In the last section we discuss the findings and implications for policy research.

## 2. BACKGROUND LITERATURE

We draw from two broad strands of literature; one is the adaptive capacity to climate risk literature and the second is information and communication technology for development (ICT4D) literature. We see that in both these areas, the role of social relations is understudied. We attempt to bridge this gap by empirically studying the linkages between the role of social relations and technology (ICTs) in enhancing farmer's capacity to adapt and being more resilient to climate shocks.

Within the adaptation literature, the broader understanding of adaptation corresponds to adjustments made by the exposed units to moderate the impacts pertaining to climate variability (IPCC, 2014, pp.838; Brooks, 2003; Ospina and Heeks, 2010). For instance, crop diversification is an adaptation practice followed by farmers as it implies change in farm practices (Billah et al., 2015). However, there is lack of agreement about the determinants of adaptive capacity across different scales viz. national, community, or household level (Jones et al., 2010).

Context specific nature of adaptation influenced by social, political, economic, and institutional factors, social identity, and power relations mediating the impact of climate hazards (Brooks and Adger, 2005) and nature of hazard itself (Brooks, 2003) which require building of specific adaptive capacity make direct measurement of adaptive capacity difficult (Brooks and Adger, 2005; Jones et al., 2010). Therefore, we take a more generic approach to understand adaptive capacity as a set of resources and system's ability and willingness to deploy these resources for achieving adaptation (Brooks and Adger, 2005). Studies include resources like natural resources, (e.g., land, water, raw materials, and biodiversity), human resources (e.g., labour, skills, knowledge and expertise), financial and economic resources, technology, social capital (e.g., strong institutions, transparent decision-making systems, formal and informal networks), institutions and networks, and equity (Brooks and Adger, 2005; Jones et al., 2010; Dillow, 2008).

Nonetheless, it is argued that tangible resources and infrastructure as determinants of adaptive capacity are given more importance while discounting the role of subjective human factors (Brown & Westaway, 2011). Although role of personal factors like entrepreneurship skills of farmers to take decisions is important, Kangogo et al. (2020) argue that interpersonal relations which can help in coping with the climate stressors require due attention. Cohen et al. (2016) argued that local social relations strongly explain the differences in the capacity to adapt or to cope with the change. Structural factors and ideology (informal institution) influence the choice of adaptation strategies that are feasible and are therefore important to take into account (Brooks & Adger, 2005). Learning outcomes through social interactions are likely to be more effective as people involved in the process share common interests and beliefs (Munasib & Jordan, 2011). Munasib & Jordan (2011) argued that associational membership promotes information sharing through increased interactions, exchange of ideas and knowledge sharing and promotes learning and informal training newer agricultural practices and thus positively influences farmer's decision to adopt sustainable agricultural practices and also the extent of such adoption.

ICTs have potential to facilitate these interactions with more frequent exchange of knowledge information among the farmers. However, the role of ICTs particularly in area of climate change adaptation is not well explored especially in global south (Ospina & Heeks, 2010). Research in this area is mainly directed, one, to understand the influence of some common demographic indicators such as age, level of education, size of the household etc. Second, towards assessing the economic gain measured in terms of increased crop productivity in terms of increased yield (Casaburi et al., 2014; Cole and Fernando, 2016), reduced production and transaction cost (Mittal, 2012), access to better market price of produce (Goyal et. al., 2010), improved access to markets (Munyua et al., 2009).

We therefore propose to incorporate the ensemble view of ICTs which allows to consider that social and contextual aspects determine how ICT is conceived (Sein & Harindranath, 2004) and to assess whether this institutional framework facilitates access to and receipt of information by the farmers. The comprehensive definition of ICT ecosystem is as proposed by Diga & May (2016) a system that encompasses the policies, strategies, processes, information, technologies, applications and stakeholders that together make up a technology environment for a country, government or an enterprise and most importantly people – diverse individuals, who create, buy, sell, regulate, manage and use technology. Integrating ICT ecosystem is important because if socio-cultural factors are influencing the use of technology, there is need to investigate how social networks that mediate the access to ICTs and receipt of information by farmers can explain differential adaptation strategies of the farmers.

# 3. METHODOLOGY

The study was conducted in one of the major wheat producing states in India viz. Haryana. Within Haryana, data was collected from six districts which were Ambala, Panchkula, Jind, Hisar, Palwal, and Mewat (see figure 1[i]). These districts were selected on the basis of three agri-climatic zones, rainfall and soil conditions. Data used in this study was collected for a randomized control trial (RCT) experimental study to study the effect of agri-met advisories sent to farmers through short message services (SMSs) on various farming outcomes. Data was collected for two consecutive rabi seasons (which starts around November and crop harvesting begins around April-May) in the year 2016-17 and 2017-18 for wheat crop. Data collected only for the baseline period 2016-17 is used in this study to avoid the bias in the intervention period data where all the farmers in treatment group received the information through SMSs.

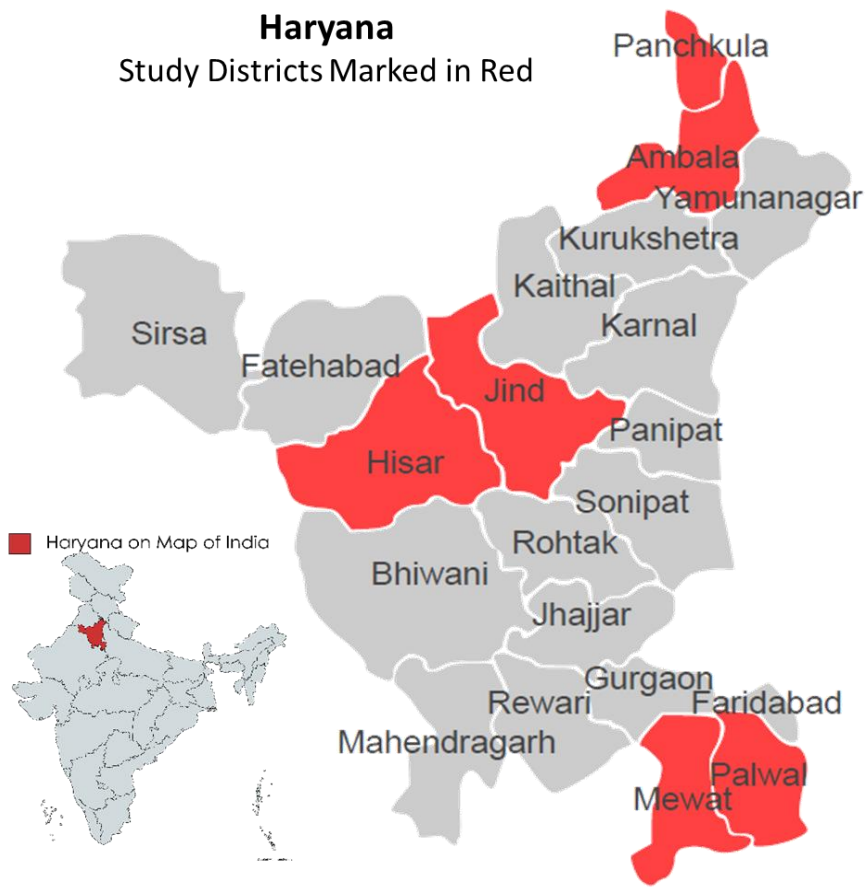

**Figure 1. Study districts on map of Haryana**

To obtain a representative sample, from each district 10 villages were selected where SMS advisories were not received by farmers. It was ensured that a pair of similar villages located at some distance to each other and are not contiguous to villages that were already receiving SMS advisories were selected so as to avoid contamination during the intervention phase. For Panchkula district, data could be collected only for 8 villages. Thus, there are total 58 villages from where data was collected. Within each village, data was collected from a minimum of 10 farmers. Keeping in mind the possibility of attrition in the future, more than 10 farmers were contacted. The initial sample size was 640 farmers which reduced to 463 farmers by the end of data collection. Initial face-to-face contact with the farmer was established in the year 2016 and subsequent data collection was done through telephonic interview. Farmer's consent was obtained before data collection. A farmer was contacted on an average 4-5 times during the entire season to collect information related to operational cost incurred on various farming operations, date of execution of such operations, ICT related information. The tool for data collection was a comprehensive structured questionnaire which was tested in the field during the pilot survey. Questionnaire was designed to collect information pertaining to cost incurred on different agricultural operations (such as land preparation, sowing, irrigation, fertilizer application, weedicides and pesticides application, harvesting, marketing, and storage) and other relevant information like timing of these operations, inputs used, credit related information etc. Information related to use of ICTs and information was also collected. Unit of analysis in the study is individual farmer.

To study the linkages between ICTs, information, and adaptive capacity of farmers, we now discuss the variables used in this study. Table 1 discusses the dimensions of adaptation decisions used in the study with rationale behind including specific indicator variables for each dimension.

| Dimension | Indicators | Description |
|---|---|---|
| Risk mitigation through crop and input diversification | Crops grown other than wheat | Works as a hedge against the vulnerability present due to single crop production facing varied risks such as undesirable changes in weather, biological risk (pest attack), market risks etc. |
|  | Grown more than one seed variety | This could help the wheat crop withstand unforeseen changes in environmental conditions. |
| Followed land management strategies | Applied organic manure | Organic manure helps in keeping the soil moisture for long and also improves health of the soil. |
|  | Used fertilizers other than DAP and urea | This indicates that farmer is investing in keeping the nutrients in the soil and not just concerned about higher crop yield. |
|  | Grown multiple crops | Growing many crops helps in balancing the nutrients in the soil and at the same time may support the crops simultaneously grown crops. |
| Taking advantage of opportunities | Utilizing rainfall | Substituting rain water for ground water irrigation saves on economic cost of expending mechanical and human labor to extract scarce water resource, enhances crop growth especially when quality of ground water is moderate, and saves from the crop loss due to excessive water. This variable demonstrates farmer's ability to take advantage of available opportunities. |
| Economic gains | Impact on yield per acre | Higher yield obtained per acre implies more revenue to the farmers which they can invest in enhancing their capacity to take more adaptation-oriented measures. |
|  | Impact on operational cost per acre | Cost is an important economic variable and lowering the cost of cultivation can support farmer economically to respond better to climate stressors. |

Note: This is a work in progress and only the first dimension is analyzed.

**Table 1. Indicators of adaptation decisions**

Based on the literature, we have conceptualized adaptive capacity as access to different resources and farmer's ability and willingness to use those resources to respond to the risk from variable weather and climatic conditions. Figure 2 depicts the conceptualization and is further elaborated in table 2 with different indicator variables.

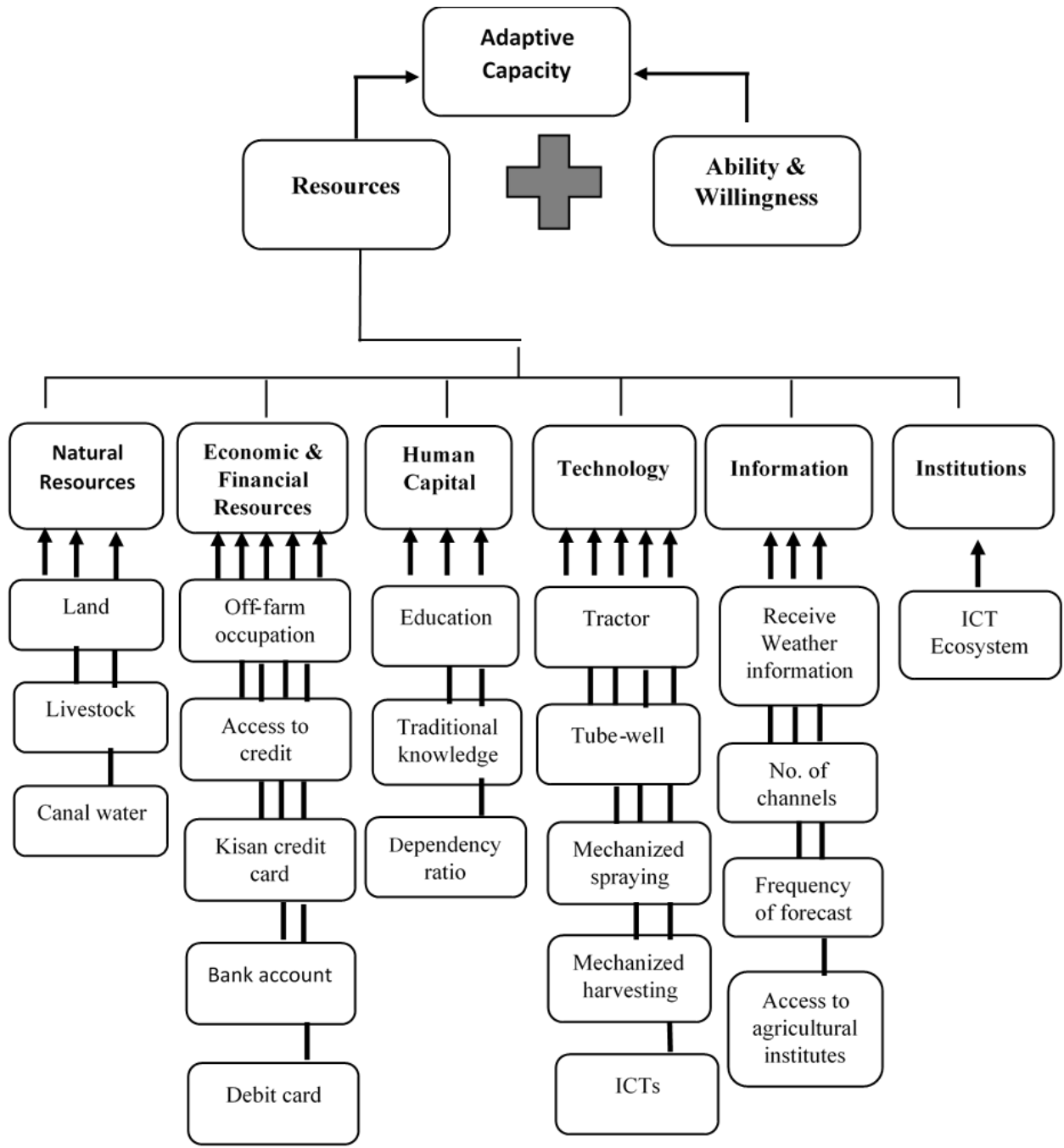

**Figure 2. Determinants of adaptive capacity (adapted from Dillow (2008))**

| Dimension | Indicators | Description |
|---|---|---|
| Natural Resources | Own livestock | Provides farmers additional source of income, inputs for the crop in the form of manure, and also provides food items such as milk and milk products, eggs etc. that helps the community to reduce reliance on farming alone. |
| | Able to use canal water | Whether a farmer was able to use canal water for irrigation (completely or partially). It is cheaper and its quality is considered as better than ground water, but available to only limited number of farmers. |
| | Land (Is not a small farmer) | Land is an important resource to a farmer. Having access to a larger piece of land provides farmer option to experiment with different crops, inputs, technologies, etc. Since all the farmers own land, we have looked at the size of operational land holding. |
| Economic and Financial Resources | Engaged in two occupations | Reduces reliance on agriculture, expands farmer's social network. |
| | Bought either seed or fertilizers on credit | Specifically captures the use of credit money in the form of working capital to purchase inputs and not for any other non-agricultural purposes. Eases liquidity for undertaking farm operations by making funds available especially during lack of cash-in-hand. |
| | Have a Kisan Credit Card (KCC) | It is a formal credit source, provides credit at a very low rate of interest to buy inputs which is often relaxed if paid back on time. |
| | Have a bank account | Represents farmer's access to formal sources of credit which provide credit at cheaper rates. Lower interest payments can lower farmer's financial burden and thereby facilitate the process of adaptation. |
| | Have a debit card | Simplifies cash withdrawal and purchase of material, saves time and efforts to access cash in bank accounts. |
| Human Capital | Education (Is literate) | Education in its own right is important for human development, further it may facilitate access to more information channels, enhanced ability to exploit technology, and adopt new methods of farming. |
| | Adult-children ratio is greater than 1 | Demonstrates availability of family labor to carry out farm operations taken in the form of whether the adult-children ratio is greater than 1 or not. |
| | Use traditional methods to guess changes in weather | Such as using environmental cues like behaviour of birds and insects, direction of wind etc. belongs to specific area and community and therefore is salient. May help farmers in taking preparatory actions. Is considered important to be combined with modern scientific knowledge. |
| Agricultural Technology (Ag_Technology) | Own tractor | Tractor is required in almost all farming activities. Ownership helps avoiding additional hiring charges and scheduling of farm operation is not hindered due to unavailability of tractor during peak timings. |
| | Own tube-well | Saves additional hiring charges, provides flexibility in operating it with diesel in case of unavailability of electricity. |

|  | Mechanized spraying | Use of tractor mounted spray pumps reduces application time and reduces dependence on human labor especially in times when quick measures are required. |
|---|---|---|
|  | Mechanized harvesting | Use of machinery instead of human labor to harvest quickens the process, shows use of available technology by farmers. |
| Information | Receive weather information | Whether a farmer simply receive weather information from any source. |
|  | Receive weather information weekly or more frequently | Captures how frequently a farmer receives weather information. |
|  | Receive weather information from more than 1 source | Captures whether the information received is from multiple or single channel; it may influence reliability and usability of information. |
|  | Contact agricultural experts through ICTs | Whether a farmer reach out to agricultural institutes for obtaining any kind of agricultural information. |
| Access to ICTs | This indicates farmers' ownership and hence access to ICT tools such as TV, smartphones, and computer with or without internet. | |

**Table 2. Indicators of Resources as Constituents of Adaptive Capacity**

Next, in table 3 we discuss the indicators used in the study to conceptualize farmers' capability and willingness to use ICTs for productive purposes particularly for agricultural work

| Dimension | Represents what? |
|---|---|
| Can and does open and read SMS | Ability to do basic functions in mobile phone and access the agri-met advisories if delivered on the phone |
| Can use online mobile applications | Ability to use more complex mobile applications |
| Mobile banking | Ability to exploit formal banking services remotely and farmer's willingness and openness to use of modern ICT channels viz. smartphone and internet to access banking services |
| Can and does use Debit card for ATM Transactions | Ability and willingness to access funds through other modes |
| Can and does use Debit card for Online Transactions |  |
| Use ICTs to search for market price of agricultural produce | These two types of uses indicate farmer's awareness about use of ICTs for other agricultural purposes (other than just accessing weather information) |
| Use ICTs to get information about government support schemes for farmers |  |
| Call Kisan Call Centre | Indicates farmer's willingness to use the available farm related advisory institutions to seek their advice |
| Seek weather information | Indicates farmer's willingness to obtain weather information for making farm decisions |

**Table 3. Indicators of Ability and Willingness of Farmers to Use ICTs**

Lastly, in table 4 we look at the indicators of farmer's ICT ecosystem that explains how ICTs are incorporated in a farmer's life and contribute to easy communication, increased interactions and exchange of knowledge and information

| Dimension | Represents what? |
|---|---|
| Use of internet for communication by family members | Depicts frequent use of modern ICT tool and family environment of the farmer with respect to preference towards newer and modern technologies. |
| Use of internet for communication among friends | This shows the pervasiveness of internet usage not only within the family but also in the friend circle of the farmer. This wider use of internet increases the likelihood to obtain varied information. |
| Use of ICTs for productive interactions with others | Interacting with input suppliers to obtain information about prices, products and other miscellaneous market and related information |
| | Connects with Agricultural universities or experts. This represent that instead of agricultural universities or expert, that are also infomediaries, reaching out to the farmers they themselves are contacting these experts through ICTs |
| | Sharing useful information with fellow farmers demonstrating the importance of close links in information dissemination |

**Table 4. Indicators of ICT Ecosystems at the Village Level**

## 4. RESULTS

We have reported the demographic profile of the farmers who participated in the present study in table 5. Majority of the farmers fall in the less than 50 years of age bracket. Average age of the farmers in the study is about 46 years. Only 17% of the farmers are illiterate. Half of the farmers have attained education till matriculate or above. Similarly, nearly half of the sample farmers belong to upper caste group. Average size of operational landholding of sample farmers is about 9 acres. The smallest farmers work on just half an acre of land. Operational landholding includes land owned and land leased in excluding the area leased out

| Farmer Characteristics | Mean/Frequency |
|---|---|
| **Age:** | |
| Mean, range, standard deviation | 45.58 years, [18-80], 13.02 |
| 35 years or less | 110 (24%) |
| 36-50 years | 195 (42%) |
| Above 50 years | 158 (34%) |
| | |
| **Education:** | |
| Illiterate & informally literate | 80 (17%) |
| Less than Primary to Middle | 152 (33%) |
| Matriculate and above | 231 (50%) |
| | |
| **Caste:** | |
| Upper Caste | 227 (49%) |
| Other Backward Caste (OBC) | 210 (45%) |
| Other Caste | 26 (6%) |
| **Land holding/Farm Size** | **Operational** |
| Mean [range] | 8.76 acres, [0.5- 62] |
| Marginal (Land holding< 2.47 acres) | 75 (16%) |
| Small (2.47 acres => Land holding< 4.95 acres) | 77 (17%) |
| Semi-Medium (4.95 acres => Land holding< 9.89 acres) | 163 (35%) |
| Medium (9.89 acres => Land holding< 24.8 acres) | 124 (27%) |
| Large (Land holding=> 24.8 acres) | 24 (5%) |

**Table 5. Description of Farmer Characteristics**

Table 6 provides the descriptive statistics of the variables discussed in the methodology section and are used in the analysis. We see that penetration of traditional ICTs like TV is quite good whereas modern ICT channels which require skills to use them still have lower ownership. Similarly, we find that most of the farmers are capable of using basic feature like opening and reading SMS and seek relevant information. However, more complex tasks such as use of technology to get information or to carry out financial transactions still scores low frequency among the sample farmers. The use of ICT channels for communication and exchange of information is more with fellow farmers and family compared to input dealers and agricultural experts. Although nearly three-quarters of the farmers reported to have received information, the other indicators of information access such as frequency and multiplicity of information sources is still low. Ownership of basic technological inputs such as tractor and tube-well is moderate.

| Variables | Response category | Mean, (SD), [Range] | Frequency (%) |
|---|---|---|---|
| **Index of Access to ICTs:** | | 1.09, (0.71), [0-3] | |
| TV ownership | Yes | | 376 (81%) |
| Owned Smartphone | Yes | | 90 (19%) |
| Owned computer with or without Internet access | Yes | | 43 (9%) |
| **Index of Capability and willingness to use ICTs:** | | 3.09, (1.60), [0-8] | |
| Can and does open and read SMS | Yes | | 402 (87%) |
| Can use online mobile applications | Yes | | 129 (28%) |
| Mobile Banking | Yes | | 30 (6%) |
| Can and does use Debit card for ATM Transactions | Yes | | 178 (38%) |
| Can and does use Debit card for Online Transactions | Yes | | 37 (8%) |
| Use ICTs to get information about government support schemes for farmers | Yes | | 88 (19%) |
| Use ICTs to search for market price of agricultural produce | Yes | | 174 (37%) |
| Call Kisan call Centre (KCC) | Yes | | 8 (2%) |
| Seek weather information | Yes | | 388 (84%) |
| **Index of ICT social ecosystem:** | | 2.24, (1.46), [0-5] | |
| Use of internet for communication by family members | Yes | | 290 (62%) |
| Use of internet for communication among friends | Yes | | 288 (62%) |
| Use of ICTs for sharing information with fellow farmers | Yes | | 242 (52%) |
| Use ICTs to contact input suppliers | Yes | | 124 (26%) |
| Use ICTs to contact agricultural experts | Yes | | 96 (20%) |
| **Index of Information Access:** | | 1.74, (1.20), [0-4] | |
| Receive weather information | Yes | | 338 (73%) |
| Receive weather information weekly or more frequently | Yes | | 209 (45%) |
| Receive weather information from more than 1 source | Yes | | 167 (36%) |
| Contact agricultural experts through ICTs | Yes | | 96 (20%) |

| **Index of Agricultural Technology used by Farmer:** | | 1.84, (1.16), [0-4] |
|---|---|---|
| Own a tractor | Yes | 218 (47%) |
| Own a tube-well | Yes | 323 (70%) |
| Mechanized weeding | Yes | 98 (21%) |
| Mechanized harvesting | Yes | 213 (46%) |
| **Index of Natural Resources:** | | 1.68, (0.75), [0-3] |
| Own livestock | Yes | 425 (92%) |
| Able to use canal water | Yes | 126 (27%) |
| Is not a small farmer | Yes | 229 (49%) |
| **Index of Economic and Financial Resources:** | | 2.56, (1.12), [0-5] |
| Engaged in two occupations (farming being primary) | Yes | 116 (25%) |
| Bought either seed or fertilizers on credit | Yes | 218 (47%) |
| Have Kisan Credit Card | Yes | 191 (41%) |
| Have a bank account | Yes | 426 (92%) |
| Have a debit card | Yes | 237 (51%) |
| **Index of Human Capital:** | | 1.92, (0.84), [0-3] |
| Is literate | Yes | 383 (83%) |
| Adult-children ratio is greater than 1 | Yes | 288 (62%) |
| Use traditional methods to guess changes in weather | Yes | 218 (47%) |
| **Risk Mitigation through Diversification** | | 0.82, (0.74), [0-2] |
| Crops grown other than wheat | Yes | 219 (47%) |
| Grown more than one seed variety | Yes | 163 (35%) |

**Table 6. Descriptive Statistics of Variables used in the Study**

However, more use of machinery for weeding and harvesting is still low among the farmers. Over 90% of the farmers own livestock which provides income and nutritional support to farming households. Nonetheless, in terms of resource endowment we see that access to canal water which is a cheaper source of irrigation is quite low and about half of the farmer have smaller piece of land to work on. Only a quarter of farmers have secondary occupation. About 50% of the farmers buy seed and fertilizers on credit. Although over 90% of the farmers have a bank account, use of plastic money or debit card is still relatively low among the farmers. Majority of the farmers are literate and about half of the farmers still use old traditional techniques to predict changes in the weather. In nearly two-third farmer households the proportion of adult members is higher than children. About half of the farmers grow only wheat as a single crop and about two-third of the farmers use only one variety of seed to grow wheat.

To better understand the linkages between information, ICTs, and farmer's adaptive capacity that contribute to building of resilience to climate stressors we use path analysis technique. Path analysis can be understood as a subset of structural equation modelling (SEM) used to estimate a system of equations. Path analysis allows us to estimate the effect of a set of variables on a specific outcome variable through multiple pathways. The advantage of this technique is that it allows to capture both direct and indirect effects of a variable by incorporating mediation analysis simultaneously. In the analysis only structural model is estimated using the observed variables with the assumption of multivariate normality. Some of the variables in our dataset do not satisfy the normality condition checked through Mardia test and univariate Q-Q plot. To account for multivariate non-normality, we have used the robust maximum likelihood estimator with Satorra-Bentler correction. The analysis was done using the lavaan package in RStudio. Figure 3 is the path diagram that depicts the relationships among the variables hypothesized and tested through path analysis. The straight-line with arrow head on one end shows the direct effect of one variable on another. The curved line with arrow head on both ends represents covariance between the two variables. The dotted lines represent that the path is not significant statistically

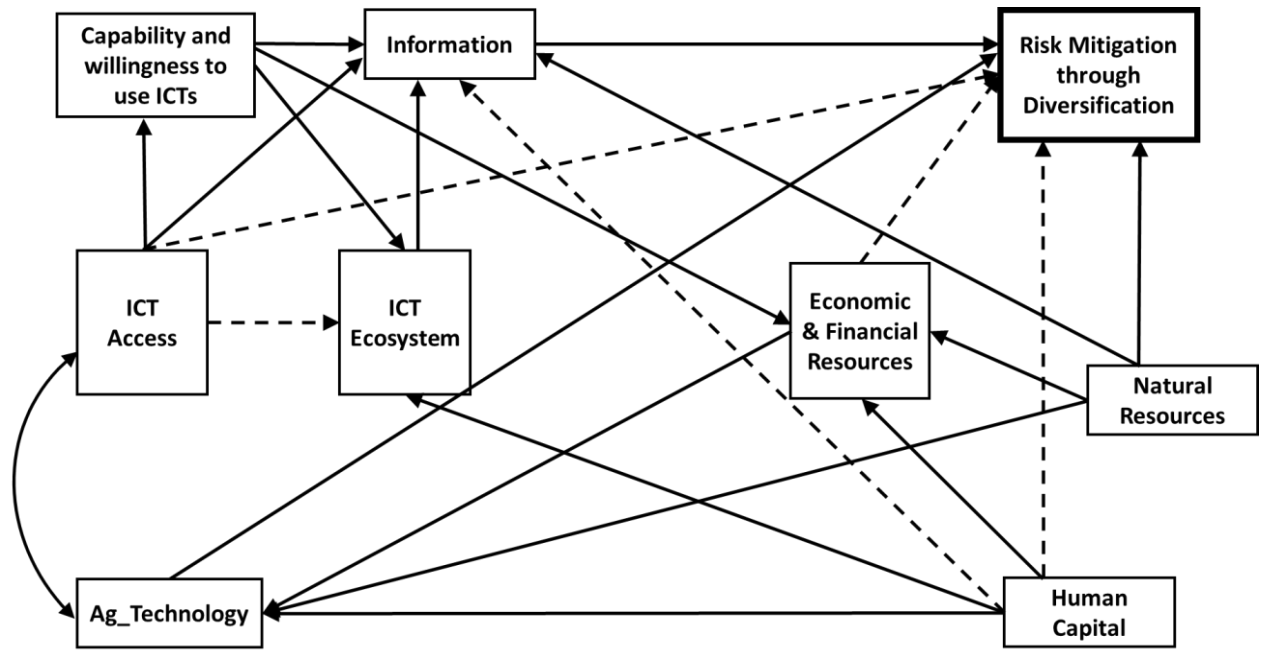

**Figure 3. Path Diagram (Direct/Indirect Effects – Standardized Solution**

Before discussing the model, it is important to check the overall fit of the model to establish its acceptance. To evaluate the model fit, we present the following fit indices in table 7 along with their criterion value and robust estimates. Based on the criterion values, the model fit is good and acceptable.

| Fit Indices | $X^{2*}$ | DF | GFI | SRMR | RMSEA | AGFI | CFI | NFI | IFI |
|---|---|---|---|---|---|---|---|---|---|
| **Criterion** | - | - | >0.90 | <0.08 | <0.08 | >0.90 | >0.90 | >0.90 | >0.90 |
| **Actual Value** | 46.92 (0.00) | 14 | 0.978 | 0.056 | 0.071 | 0.930 | 0.930 | 0.907 | 0.933 |
| **Robust Value** | 48.25 (0.00) | 14 | - | 0.056 | 0.073 | - | 0.929 | - | - |

Note: * indicates the Satorra-Bentler chi-square value. Values in brackets indicate p-value
**Table 7. Model Fit Indices**

Table 8 presents the standardized path coefficients. The level of significance guides whether the proposed hypotheses are supported or not as reported in the inference column.

| Path | Standardized Path Coefficients | Robust Std. Error | P-Value | Inference |
|---|---|---|---|---|
| Information → Risk_Mitigation | 0.100** | 0.028 | 0.028 | Supported |
| Econ_Fin_Resources → Risk_Mitigation | 0.027 | 0.031 | 0.570 | Not Supported |
| Ag_Technology → Risk_Mitigation | 0.154*** | 0.032 | 0.002 | Supported |
| Human_Capital → Risk_Mitigation | -0.009 | 0.042 | 0.844 | Not Supported |
| Natural_Resources → Risk_Mitigation | 0.178*** | 0.047 | 0.000 | Supported |
| ICT_Access → Risk_Mitigation | -0.035 | 0.052 | 0.477 | Not Supported |
| ICT_Ecosystem → Information | 0.156*** | 0.041 | 0.002 | Supported |
| Natural_Resources → Information | 0.078* | 0.069 | 0.070 | Supported |
| Human_Capital → Information | 0.039 | 0.063 | 0.379 | Not Supported |
| ICT_Capability_Willingness → Information | 0.105** | 0.039 | 0.040 | Supported |
| ICT_Access → Information | 0.122*** | 0.077 | 0.008 | Supported |
| Natural_Resources → Ag_Technology | 0.280*** | 0.065 | 0.000 | Supported |
| Econ_Fin_Resources → Ag_Technology | 0.107*** | 0.042 | 0.008 | Supported |
| Human_Capital → Ag_Technology | 0.095** | 0.055 | 0.020 | Supported |
| ICT_Capability_Willingness → Econ_Fin_Resources | 0.292*** | 0.028 | 0.000 | Supported |
| Natural_Resources → Econ_Fin_Resources | 0.173*** | 0.069 | 0.000 | Supported |
| Human_Capital → Econ_Fin_Resources | 0.153*** | 0.055 | 0.000 | Supported |
| Human_Capital → ICT_Ecosystem | -0.126*** | 0.068 | 0.001 | Supported |
| ICT_Capability_Willingness → ICT_Ecosystem | 0.486*** | 0.037 | 0.000 | Supported |
| ICT_Access → ICT_Ecosystem | 0.030 | 0.079 | 0.435 | Not Supported |
| ICT_Access → ICT_Capability_Willingness | 0.346*** | 0.101 | 0.000 | Supported |

Note: *** indicates a significance level of 1%, ** indicates a significance level of 5%, and * indicates a significance level of 10%.

**Table 8. Path Coefficients**

We find that information is a significant contributor to farmer's adaptive capacity. Farmers who have higher information index, indicating better information access, diversify their risk more (0.100). It is likely as getting relevant information particularly from a greater number of sources with higher frequency of receiving information and when expert's advice is involved may help farmers to use variety of seeds and crops that works as hedge against the risk of crop failure. Ownership of technological inputs like tractor and tube-well and use of machinery to carry out agricultural operations makes it easier for farmers to diversify (0.154).

Natural resources also play significant role in enhancing farmer's adaptive capacity by helping them to mitigate the risk through diversification (0.178). Having greater endowment of natural resources particularly land facilitate diversification. However, these results also indicate the likelihood of greater capacity to adapt for those who are richer or have higher resource endowment. As we argued at the beginning of the study, though other resources are necessary to undertake adaptation decisions, augmenting these resources especially in the short-term is a difficult task especially for the resource poor farmer. The significant linkages among capability and willingness to use ICTs for productive purposes, ICT ecosystem, and information implies that there is not only direct effect of information but there are other interlinked mechanisms that influence availability and use of information by the farmers and hence their capacity to adapt.

Indirect effect is the effect of one variable on another mediated by a third variable called mediating variable. Table 9 presents the indirect effects pertaining to dimensions of ICTs, information, and risk mitigation. In table 9 we see that farmers' embeddedness into the ICT ecosystem significantly influence their ability to respond to risk by facilitating their access to information (IE1 in table 9). While we see that farmers' capability and willingness to use ICTs for productive purposes play important role in ensuring information access directly (table 8), its indirect effect is, however, not significant in risk mitigation through this channel (IE2 in table 9). On the other hand, when this linkage is mediated by ICT ecosystem, we see that capability and willingness also contribute to risk mitigation (IE3 in table 9). The role of ICTs in terms of having only physical access only is facilitating access to information (IE5 in table 9) and does not directly helps in risk mitigation (table 8). The indirect effects mediated through information suggest that enhancing farmer's access to information can work as a catalyst for enhancing adaptive capacity of farmers. Thus, the role of information and ICT ecosystem that promotes exchange of information especially agricultural information becomes even more significant.

| **Indirect Effect (IE)** | **Standardized Estimate** | **P-value** |
|---|---|---|
| **IE1:** ICT_Ecosystem → Information → Risk_Mitigation | 0.016* | 0.065 |
| **IE2:** ICT_Capability_willingness → Information → Risk_Mitigation | 0.011 | 0.127 |
| **IE3:** ICT_Capability_willingness → ICT_Ecosystem → Information → Risk_Mitigation | 0.008* | 0.067 |
| **IE4:** ICT_Access → ICT_Capability_willingness → Information → Risk_Mitigation | 0.004 | 0.136 |
| **IE5:** ICT_Access → Information → Risk_Mitigation | 0.012* | 0.091 |

Note: *** indicates a significance level of 1%, ** indicates a significance level of 5%, and * indicates a significance level of 10%

**Table 9. Indirect Effects**

Thus, the present study not only propose to invest in skills to utilize available channels of information particularly ICTs but also draws attention to take a more systemic perspective to incorporate the use of ICTs within the farming community to enhance the adaptive capacity of farmers and hence make them more resilient to the climate stressors. Although developing a facilitating ICT ecosystem for the farmers requires time and dedicated efforts to understand the nuances of local social settings, our study suggests that it can help farmers to make important agricultural decisions even with the limited resources at their disposal.

## 5. DISCUSSION

The present study reveals that information is an important constituent of farmers' adaptive capacity to climate risk. However, the linkages capability and willingness to use ICTs to access and use information and also the exchange of knowledge and information facilitated by ICT ecosystem are important to be exploited by any intervention that seek to facilitate adaptation through modern days technology (ICTs and IoTs: internet of things). In order to understand association between technology and social relations, it is important to understand how the introduction and use of technologies empower networked people vis-à-vis those who are not connected across different class, caste, gender and regions. For instance, Ali & Kumar (2011) discussed, even though information delivery through ICTs led to improvement in the quality of decision making of user farmers vis-à-vis non-user farmers, user farmers belonging to socially lower class had no difference in decision making compared to non-users of the same social class. We argue that adopting a more systemic approach allows to go beyond the access-capability enhancement dyad in the ICT domain and take a more holistic view of access and use of information while looking at the engagement between social relations and technology.

Learning outcomes through social interactions are likely to be more effective as people involved in the process share common interests and beliefs (Munasib & Jordan, 2011). Thus, strengthening local ICT ecosystem of farmers could be a more sustainable and welcomed by the famers. Encouraging farmers to participate in an ICT ecosystem may give them a sense of belongingness and therefore is more likely to be embraced by the farmers increasing their likelihood to be more adaptive and resilient to climate stress

---

[i] Maps were created using the following sources: https://gramener.com/indiamap/, https://mapchart.net/india.html